\documentclass[twoside,12pt]{article}
\usepackage{epsfig}

\def\Journal#1#2#3#4{{#1} {#2} (#4) #3 }

\def\NPA{{\em Nucl. Phys.} A}

\def\NPB{{\em Nucl. Phys.} B}

\def\PLB{{\em Phys. Lett.} B}

\def\PRL{\em Phys. Rev. Lett.}

\def\PREP{\em Phys. Rep.}

\def\PRD{{\em Phys. Rev.} D}

\newcommand{\be}{\begin{equation}}
\newcommand{\ee}{\end{equation}}
\newcommand{\bea}{\begin{eqnarray}}
\newcommand{\eea}{\end{eqnarray}}

\begin{document}

\title{ \vspace{1cm} Gluons, tadpoles, and color neutrality 
in a two-flavor color superconductor}
\author{Dennis D.\ Dietrich$^{1}$ and Dirk H.\ Rischke$^2$ \\
\\
$^1$Laboratoire de Physique Th\'eorique, Universit\'e  Paris XI, 
Orsay, France\\
$^2$Institute for Theoretical Physics, University Frankfurt, Germany}
\maketitle
\begin{abstract} Considering cold, dense quark matter with two massless
quark flavors, we demonstrate how, in a self-consistent calculation
in the framework of QCD,
the condensation of Cooper pairs induces a non-vanishing background
color field. This background color field has precisely
the right magnitude to cancel tadpole contributions and thus
ensures overall color neutrality of the two-flavor color superconductor.
\end{abstract}
\section{Introduction and Conclusions} \label{I}

Due to asymptotic freedom \cite{asym}, at quark chemical
potentials $\mu \gg \Lambda_{\rm QCD}$ single-gluon exchange is the
dominant interaction between quarks in cold, dense quark matter. 
Single-gluon exchange is attractive in the color-antitriplet channel, which
gives rise to the formation of quark Cooper pairs \cite{CSC}.
As this is analogous to what happens in ordinary superconductors
\cite{BCS}, this phenomenon was termed {\em color superconductivity\/} 
\cite{Reviews,RischkeReview}.

A well-studied color-superconducting system is cold, dense quark matter
with two flavors of massless quarks \cite{Reviews,RischkeReview}. 
In this case, up and down quarks of, say, red and green color
form anti-blue Cooper pairs with total spin zero in the color-antitriplet, 
flavor-singlet channel.
Blue up and down quarks remain unpaired.

If there is a single chemical potential $\mu_u = \mu_d \equiv \mu$ 
for quark number, the system is color-neutral in the normal-conducting
phase, and the number densities of red, green, and blue quarks are
equal,
\be
n_r^0 = n_g^0 = n_b^0 \equiv 2 N_f \int \frac{{\rm d}^3 {\bf q}}{(2 \pi)^3}\,
n_{\bf q}^0\;,
\ee
where $n_{\bf q}^0 \equiv \Theta(\mu-q)$
is the occupation number for non-interacting, massless quarks.
A natural question is then whether such a system is still color-neutral 
in the superconducting phase. Naively, one would think that
this should be the case, as the total number of quarks does not change
just because a part of them has formed Cooper pairs.
However, at fixed quark chemical potential $\mu$ this is in fact not true, 
because pairing changes the
dispersion relation for quasi-particle excitations \cite{RDPDHRsuperfluid},
\be
\epsilon_{\bf q}^{0} \equiv |\mu - q | \; \rightarrow \; 
\epsilon_{\bf q} \equiv \sqrt{(\mu - q)^2 + \phi^2}\;,
\ee
where $\phi$ is the color-superconducting gap function.
In turn, also the
occupation number for quasi-particle excitations changes 
\cite{RDPDHRsuperfluid},
\be \label{occ}
n_{\bf q}^0 = \Theta(\mu - q) \; \rightarrow \; n_{\bf q} = 
\frac{\epsilon_{\bf q} + \mu - q}{2\, \epsilon_{\bf q}}\;.
\ee
It is not hard to see that, due to phase space ${\rm d}^3 {\bf q}
\sim {\rm d}q \, q^2$ and the symmetric ``smearing'' of the quasi-particle
occupation number (\ref{occ}) around the Fermi surface, 
in a system with fixed quark chemical potential $\mu$ the
total density of quasi-particle excitations corresponding to
paired red and green quarks is
{\em larger\/} than that for unpaired red, green, and blue quarks,
\be \label{nrg}
n_r = n_g \equiv 2 N_f \int \frac{{\rm d}^3{\bf q}}{(2\pi)^3} \, n_{\bf q}
> n^0_r = n^0_g= n_b^0  \;.
\ee
This mismatch in color causes the system to be no longer color-neutral;
it carries an excess of anti-blue (i.e., red and green in the
color-antitriplet channel) over blue color charge.
The integral in Eq.\ (\ref{nrg}) appears to be ultraviolet
divergent, but is actually not, because the gap function $\phi$ 
vanishes rapidly away from the Fermi surface, see also Sec.\ \ref{IV}.

It is a well-known fact that, after fixing the gauge, the formation
of Cooper pairs in a two-flavor color superconductor {\em spontaneously\/}
breaks $SU(3)_c$ to $SU(2)_c$. This remaining local gauge symmetry 
acts in the space of red and green quarks and is indeed preserved
in the superconducting phase, because
the number of red and green quasi-particle excitations is equal,
cf.\ Eq.\ (\ref{nrg}). However, according to the above argument, 
at fixed quark chemical potential $\mu$ the $SU(3)_c$ gauge symmetry appears
also to be
{\em explicitly\/} broken by the excess of anti-blue over blue color charge.
Nevertheless, {\em the system as a whole should still be color-neutral}.
The question then is how color neutrality can be restored.
One possibility is to introduce a color-chemical potential
$\mu_8$ ``by hand'', which reduces the Fermi surface for red and green
quarks and enlarges the Fermi surface for blue quarks in a way
to ensure color neutrality. This is the accepted procedure if one studies
color superconductivity in the framework of Nambu--Jona-Lasinio-type
models \cite{NJL,SBRDHR}. In QCD, however, one would think
that a more elegant solution is realized.

In Ref.\ \cite{Rebhan}, Gerhold and Rebhan argued that 
the inequality of red and green quasi-particle excitations
compared to the number of unpaired blue quarks induces
non-vanishing tadpole contributions, see Eq.\ (25) of Ref.\ \cite{Rebhan}:
\be \label{Jmua}
{\cal T}^{\mu}_a
\equiv \frac{2\,g}{ \pi^2}\,\mu\, \phi^2 \, \ln \left( \frac{2\,\mu}{\phi}
\right)\, (T_a)_{33} \, g^{\mu0}
= - \sqrt{6} \, \mu \, \phi^2 \, \delta_{a8} \, g^{\mu 0}\;.
\ee
Here, $g$ is the strong coupling constant and
$(T_a)_{33}$ is the (33)-component of the generator $T_a$ of 
$SU(3)_c$.
The last equality follows from
the leading-order result $\phi \simeq 2 \mu \exp[-3\pi^2/(\sqrt{2}g)]$
for the color-superconducting gap parameter 
\cite{gapequation} and from
$(T_a)_{33}\equiv - \delta_{a8}/\sqrt{3}$. 
(For the definition of the tadpole ${\cal T}^\mu_a$, see
Ref.\ \cite{Rebhan}, Eq.\ (15), or Eq.\ (\ref{tadpole}) below.)
It was argued in
Ref.\ \cite{Rebhan} that this tadpole contribution gives rise
to an effective color-chemical potential associated with
the eighth generator $T_8$ of $SU(3)_c$,
\be \label{mu8}
\mu_8 = -  g \, \frac{1}{3 m_g^2} \,\sqrt{6}\,  \mu \, \phi^2 
= - \sqrt{6} \,  \pi^2 \,\frac{\phi^2}{g \mu}\;,
\ee
where $m_g \equiv g\mu/(\sqrt{2} \pi)$ is the gluon mass
parameter in cold, dense quark matter with $N_f = 2$ flavors
of massless quarks.
The prefactor arises from attaching a static, electric gluon with
propagator $1/(3m_g^2)$ to the tadpole (\ref{Jmua}), with
coupling constant $g$. The resulting
diagram then affects the quark propagator in the
manner of a chemical potential.
Moreover, Gerhold and Rebhan concluded that the total
system is color-neutral, because in a gauge which does not involve
the zeroth component of the gauge field, $A_0^a$, the generating functional
contains a functional integral
$\int {\cal D} A_0^a \exp [ i \int_X A_0^a N_a ] \equiv \delta[N_a]$, 
where $N_a$ is the density of adjoint color charge $a$. The 
$\delta$-functional then ensures that $N_a \equiv 0$.

The present paper can be considered as an addendum to Ref.\ \cite{Rebhan}
and is motivated by the following argument.
In general, superconductivity is a non-perturbative phenomenon.
A perturbative calculation will never produce a gap in the
quasi-particle excitation spectrum;
the gap parameter has to be determined 
from a self-consistent solution (possibly within a certain many-body
approximation scheme) of a Dyson-Schwinger equation for the
quark two-point function \cite{BCS}. (Usually, one employs
the mean-field approximation to compute the gap, for more
details, see Refs.\ \cite{RischkeReview,SBRDHR} or
Secs.\ \ref{II} and \ref{III} below.)
However, if the gap is computed in a self-consistent manner,
then one should also incorporate the 
color-chemical potential (\ref{mu8}) in a self-consistent way.

This is done in the present paper. Before we go into
the details of the calculation, let us briefly state our conclusions.
We shall indeed obtain an effective chemical potential $\mu_8$,
which precisely assumes the value (\ref{mu8}). 
In a self-consistent calculation, this chemical potential 
does not arise from a nonvanishing tadpole (\ref{Jmua}), 
but from a {\em non-vanishing expectation value of the
gluon field}, $A_\mu^a \equiv A\, g_{\mu 0} \, \delta^{a8}$,
with $A = const.$. This background field is a self-consistent solution of
the Yang-Mills equation and acts like 
an effective chemical potential, $\mu_8 \equiv g A$, in the
quark propagator. The possibility of a non-vanishing expectation
value for the gluon field was also mentioned in Ref.\ \cite{Rebhan}.
Then, computing the tadpole self-consistently with this background field,
we shall see that it has precisely the right magnitude to
{\em cancel\/} the tadpole (\ref{Jmua}). This cancellation
is of crucial importance, as the tadpole (\ref{Jmua}) is, up
to a factor of $g$, just the color-charge
density associated with the eighth direction in adjoint color space.
The vanishing of the tadpole due to the non-zero color background
field thus ensures color neutrality in
the self-consistent calculation.

The remainder of this paper is organized as follows. In Sec.\ \ref{II}
we introduce the effective action 
for color-superconducting quark matter. To this end, it is most convenient to
work in the framework of the Cornwall-Jackiw-Tomboulis (CJT) formalism 
\cite{CJT}. In Sec.\ \ref{III} we
derive and discuss the stationarity conditions for the effective
action which determine the expectation values for the one- and
two-point functions of the theory. The former are the Dirac equation
for the quark field and the Yang-Mills equation for the gluon field,
while the latter represent Dyson-Schwinger equations for the quark
and gluon propagator. In Sec.\ \ref{IV} we solve the Yang-Mills
equation self-consistently and show that the solution is
a non-zero expectation value for the gluon field which cancels
the tadpole contribution.
Finally, in Sec.\ \ref{V} we demonstrate how this non-zero
background field provides color neutrality in the color superconductor.

Our units are $\hbar=c=k_B=1$. Four-vectors are denoted by capital
letters, $K^\mu \equiv (k_0, {\bf k})$, where
the three-vector ${\bf k}$ has modulus $k \equiv |{\bf k}|$ and
direction $\hat{\bf k}\equiv {\bf k}/k$. We work in Euclidean space
with imaginary time $\tau \equiv i\, t = i x_0$, but nevertheless
adhere to a notation familiar from Minkowski space, i.e.,
$K^\mu \equiv (k_0,{\bf k}) \equiv g^{\mu \nu} K_\nu$,
with the metric tensor $g^{\mu\nu} = {\rm diag}(+,-,-,-)$.
Space-time integrals are defined as $\int_X \equiv  \int_0^{1/T} 
{\rm d}\tau \int_V {\rm d}^3{\bf x}$, while the four-dimensional
$\delta$-function is $\delta^{(4)}(X-Y) \equiv -i \,\delta(x_0-y_0)
\delta^{(3)}({\bf x} - {\bf y})$. Integrals in momentum space
are defined as $\int_K \equiv T \sum_n \int {\rm d}^3 {\bf k}/(2 \pi)^3$,
where the sum runs over the Matsubara frequencies $\omega_n^{\rm b} 
= 2n \pi T$ for bosons and $\omega_n^{\rm f}= (2n+1) \pi T$ for 
fermions, $n = 0, \pm 1, \pm 2, \ldots$.

\section{Effective action for color-superconducting quark matter}
\label{II}

In superconducting systems, it is advantageous 
to introduce charge-conjugate fermion degrees of freedom in addition to the
``usual'' fermions and work in the so-called Nambu-Gor'kov basis \cite{BCS}.
Quarks with $N_c$ colors and $N_f$ flavors are
then described by the $8N_cN_f$-component spinors \cite{RDPDHRsuperfluid}
\be \label{NGbasis}
\Psi \equiv \left( \begin{array}{c} \psi \\ 
                                    \psi_C
                   \end{array} \right)\, , \;\;
\bar{\Psi} \equiv \left( \bar{\psi}, \bar{\psi}_C \right) \; ,
\ee
where $\psi$ is the usual $4N_c N_f$-component quark spinor, 
$\bar{\psi} \equiv \psi^\dagger \gamma_0$ the adjoint quark spinor,
and $\psi_C \equiv C \bar{\psi}^T$ the charge-conjugate quark spinor,
with the adjoint $\bar{\psi}_C \equiv \psi^T C$. Here, $C \equiv
i \gamma^2 \gamma_0$ is the charge conjugation matrix, $C^{-1} \equiv C^T
\equiv C^\dagger \equiv - C$. The QCD tree-level action 
reads \cite{RischkeReview}
\be \label{action}
I[\bar{\Psi},\Psi,A] = - \frac{1}{4} \int_X  F^{\mu \nu}_a(X) \,
F_{\mu \nu}^a(X) 
+ \frac{1}{2} \int_{X,Y} \bar{\Psi}(X) \, S_0^{-1}(X,Y)\, \Psi(Y)\;,
\ee
where $F_{\mu \nu}^a \equiv \partial_\mu A_\nu^a - \partial_\nu A_\mu^a
+ g f^{abc} A_\mu^b A_\nu^c$ is the gluon field strength tensor.
The factor $1/2$ in front of the fermionic contribution compensates
for the doubling of the degrees of freedom in the Nambu-Gor'kov basis
(\ref{NGbasis}). The inverse tree-level quark propagator in the presence of
a gluon field $A_\mu^a$ and a quark number chemical potential $\mu$ is 
\be 
\label{treequarkprop}
S_0^{-1}(X,Y) \equiv - 2 \frac{\delta^2 I [\bar{\Psi},\Psi,A]}{\delta 
             \bar{\Psi}(X) \, \delta \Psi(Y)}
          = \left( \begin{array}{cc} 
             i \gamma_\mu D^\mu_X + \mu \gamma_0 -m & 0 \\
             0 & i \gamma_\mu D^{\mu}_{C, X} - \mu \gamma_0 -m
                       \end{array} \right) \,\delta^{(4)}(X-Y)\;.
\ee
The dependence on the gluon field enters through
the covariant derivative $D^\mu_X \equiv \partial^\mu_X
- ig A^\mu_a(X) T_a$ and its charge-conjugate counter-part
$D^\mu_{C, X} \equiv \partial^\mu_X + ig A^\mu_a(X) T_a^T$.
As we shall show below, in a two-flavor
color superconductor \cite{RischkeReview}, $A_\mu^a$ assumes a classical 
expectation value $A_\mu^a \equiv A\, g_{\mu 0} \,\delta^{a8}$.
(An analogous conclusion for the so-called color-flavor-locked phase
\cite{CFL} was drawn in Ref.\ \cite{Kryjevski}).
From Eq.\ (\ref{treequarkprop}) it is obvious that
this expectation value assumes the role of a color-chemical
potential $\mu_8\equiv g A$ associated with the color-charge
generator $T_8$ \cite{SBRDHR}.

The manner in which Nambu-Gor'kov quark spinors couple to the gluon
field in Eq.\ (\ref{treequarkprop}) suggests the following definition
for the quark-gluon vertex in Nambu-Gor'kov space,
\be \label{NGvertex}
  \Gamma_a^\mu=\left(\begin{array}{cc}
    \gamma^\mu T_a & 0 \\
    0 & -\gamma^\mu T_a^T
  \end{array}\right)\;,
\ee
such that the quark-gluon coupling in the action (\ref{action})
can be simply written as $g\, A_\mu^a \, \bar{\Psi} \Gamma_a^\mu \Psi$.

The most convenient way to derive the gap equations for
color-superconducting quark matter is within the framework
of the CJT formalism \cite{CJT} where
the effective action of QCD takes the form
\cite{RischkeReview,SBRDHR,Abuki}
\bea
  \Gamma\left[\bar{\Psi},\Psi,A,S,\Delta\right]
  &=&I\left[\bar{\Psi},\Psi,A\right]
  -\frac{1}{2}\, {\rm Tr}\ln \Delta^{-1}
  -\frac{1}{2}\, {\rm Tr} \left(\Delta_0^{-1}\Delta-1\right)\nonumber\\
  &+&\frac{1}{2} \, {\rm Tr}\ln S^{-1}
  +\frac{1}{2}\, {\rm Tr} \left(S_0^{-1}S-1\right)
  +\Gamma_2\left[\bar{\Psi},\Psi,A,S,\Delta\right]\; .
  \label{Gamma}
\eea
From now on, $\bar{\Psi},\, \Psi,\, A$ denote the {\em expectation values\/}
of the quark and gluon fields.
The quantities $\Delta$ and $S$ are the full gluon and quark propagators,
respectively. The inverse tree-level quark propagator $S_0^{-1}$ was
already introduced in Eq.\ (\ref{treequarkprop}). Its gluonic counter-part
$\Delta_0^{-1}$ can be computed from Eq.\ (\ref{action}) via
\be \label{treegluonprop}
{\Delta_0^{-1}}^{\mu \nu}_{ab} (X,Y) \equiv \frac{\delta^2 I[\bar{\Psi},
\Psi,A]}{\delta A_\mu ^a(X) \, \delta A_\nu^b (Y)}\;,
\ee
but we refrain from giving its explicit form, since it will not be
required in the following. However, we note that, similar to the inverse 
tree-level quark propagator (\ref{treequarkprop}), 
also the inverse tree-level gluon propagator depends on 
the gluon field $A_\mu^a$.

The traces in Eq.\ (\ref{Gamma})
run over space-time, Nambu-Gor'kov, color, flavor, and 
Dirac indices. The functional $\Gamma_2$ is the sum of all two-particle 
irreducible (2PI) diagrams. It is impossible to evaluate all 2PI diagrams 
exactly. However, the advantage of the CJT effective action 
(\ref{Gamma}) is that truncating the sum $\Gamma_2$ after a finite 
number of terms still provides a well-defined many-body approximation. 
Here we only include the sunset-type diagram shown in Fig.\ 1 of
Ref.\ \cite{SBRDHR}, which leads to the so-called mean-field approximation,
\be  \label{Gamma2}
  \Gamma_2=-\,\frac{g^2}{4}\int_{X,Y}
  {\rm Tr} \left[\Gamma_a^\mu\,S(X,Y)\,\Gamma_b^\nu\,S(Y,X)\right]\,
  \Delta_{\mu\nu}^{ab}(X,Y)\;,
\ee
where the trace now runs only over Nambu-Gor'kov, color, flavor, 
and Dirac indices, and $\Gamma_a^\mu,\, \Gamma_b^\nu$ are the Nambu-Gor'kov
vertices introduced in Eq.\ (\ref{NGvertex}).

So far, we have not discussed the question of gauge fixing.
This can be done on the level of the inverse tree-level gluon 
propagator (\ref{treegluonprop}), where appropriate terms
have to be added.
One also has to include ghost degrees of freedom
in the CJT effective action (\ref{Gamma}). In principle, these are necessary
to cancel the contribution from the unphysical gluon degrees of freedom
to the effective action,
but at temperatures of relevance for color superconductivity,
$T \sim \phi \sim \mu \exp(-1/g) \ll \mu$ in weak coupling,
they can be neglected \cite{RischkeReview}.

\section{Stationarity conditions for the effective action} \label{III}

The CJT formalism allows to compute the expectation values for 
one- and two-point functions of the theory from the stationarity
conditions
\bea
0 & = & \frac{\delta \Gamma}{\delta \bar{\Psi}} = 
\frac{\delta \Gamma}{\delta \Psi} \; , \label{Dirac} \\
0 & = & \frac{\delta \Gamma}{\delta A} \; , \label{YM} \\
0 & = & \frac{\delta \Gamma}{\delta S} \; , \label{DSEquark} \\
0 & = & \frac{\delta \Gamma}{\delta \Delta} \;. \label{DSEgluon}
\eea
Performing the variation of $\Gamma$ with respect to $\bar{\Psi}$,
Eq.\ (\ref{Dirac}) yields the Dirac equation 
for the Nambu-Gor'kov quark spinor $\Psi$ 
in an external gluon field (correspondingly,
the variation with respect to $\Psi$ yields the Dirac
equation for the adjoint spinor $\bar{\Psi}$). We do not
give this equation explicitly, as the
solution {\em has\/} to be trivial, $\Psi \equiv 0$ (and, correspondingly,
$\bar{\Psi} = 0$), since Grassmann-valued
fields cannot have a classical expectation value.

Equation (\ref{YM}) is the Yang-Mills equation for the gluon field,
\be \label{YM2}
{\cal D}_{X\nu}^{ab}\, F^{\nu\mu}_b(X)
  =\frac{\delta}{\delta A_\mu^a(X)}\,
  \left[\frac{1}{2}\,{\rm Tr}
  \left(\Delta_0^{-1} \Delta - S_0^{-1} S\right)-\Gamma_2\right]\;,
\ee
where ${\cal D}_{X\nu}^{ab}=\partial_{X\nu} \delta^{ab}-
g f^{abc} A_\nu^c(X)$ is the covariant derivative in the 
adjoint representation. The first two terms on the right-hand side are 
the contributions from gluon and quark tadpoles.
We should add a remark concerning our nomenclature.
The first term involves a gluon loop, wherefore we choose to call 
it the ``gluon tadpole'' contribution, while the second involves
a quark loop (also recognizable by the relative minus sign with
respect to the gluon tadpole), wherefore we call it the
``quark tadpole''.
This nomenclature is based on the ``body'' of the tadpole, which
can consist of a gluon or a quark loop. It should not be confused with that of 
Ref.\ \cite{Rebhan} where tadpoles are named according
to their ``tail'', i.e., they are always referred to as ``gluon'' tadpoles.
The functional derivative with respect to $A_\mu^a$ acting on the 
trace is non-trivial because of the dependence of the 
inverse tree-level propagators $\Delta_0^{-1}$ and $S_0^{-1}$ on 
the gluon field, cf.\ Eq.\ (\ref{treequarkprop}). 
The last term is non-zero, if $\Gamma_2$ contains 2PI diagrams with an 
explicit dependence on $A_\mu^a$. It vanishes in our approximation 
(\ref{Gamma2}) for $\Gamma_2$. We solve the Yang-Mills equation (\ref{YM2})
explicitly in Sec.\ \ref{IV}.

Equation (\ref{DSEquark}) is a Dyson-Schwinger equation
for the full quark propagator,
\be \label{DSE2}
S^{-1}(X,Y) = S_0^{-1}(X,Y) + \Sigma(X,Y)\;,
\ee
where the quark self-energy is defined as
\be \label{Sigma}
\Sigma(X,Y) \equiv 2 \, \frac{\delta \Gamma_2}{\delta S(Y,X)} \;.
\ee
In the mean-field approximation (\ref{Gamma2}) for $\Gamma_2$, 
\be
\Sigma(X,Y) = - g^2 \Gamma^\mu_a\, S(X,Y) \, \Gamma^\nu_b \, 
\Delta_{\mu \nu}^{ab}(X,Y)\;.
\ee
Assuming translational invariance, 
this becomes the usual result in momentum space \cite{RischkeReview},
\be
\Sigma(K) = - g^2 \int_Q 
\Gamma^\mu_a\, S(Q) \, \Gamma^\nu_b \, \Delta_{\mu \nu}^{ab}(K-Q)\;.
\ee
As discussed in great detail in Ref.\ \cite{RischkeReview}, the
diagonal Nambu-Gor'kov components of this expression correspond to the
one-loop self-energy for quarks and charge-conjugate
quarks, respectively. The off-diagonal components correspond
to the gap equation for the color-superconducting gap parameter,
which has been solved in Ref.\ \cite{gapequation}.

However, {\em all\/} previous solutions of the
QCD gap equation have neglected the possibility
that the full quark propagator $S$ may (implicitly) depend on 
a non-vanishing expectation value for the gluon field.
This dependence of $S$ on $A_\mu^a$ enters via Eq.\ (\ref{DSE2})
through the dependence of the tree-level
propagator $S_0^{-1}$, Eq.\ (\ref{treequarkprop}), on $A_\mu^a$.
As will be shown below, the numerical value for the 
expectation value of the gluon field is of order 
$\phi^2/(g^2\mu)$. This is parametrically
of higher order, such that the previous solution of the QCD gap 
equation, which was obtained to subleading order in the strong
coupling constant $g$, remains correct.
(For a definition of the term ``subleading order'' in the context 
of the QCD gap equation, see Ref.\ \cite{RischkeReview}.)
Nevertheless, the non-zero expectation value of the gluon field
will prove to be important to ensure color neutrality
in the color superconductor.

Finally, Eq.\ (\ref{DSEgluon}) is a Dyson-Schwinger equation for
the gluon propagator,
\be
{\Delta^{-1}}^{\mu \nu}_{ab} (X,Y) = {\Delta^{-1}_0}^{\mu \nu}_{ab} (X,Y)
+ \Pi^{\mu \nu}_{ab}(X,Y)\;.
\ee
The gluon self-energy is defined as
\be
\Pi^{\mu \nu}_{ab}(X,Y) \equiv -  2 \, 
\frac{\delta \Gamma_2}{\delta \Delta_{\nu \mu}^{ba}(Y,X)}\;,
\ee
which, in the mean-field approximation (\ref{Gamma2}), becomes
\be
\Pi^{\mu \nu}_{ab}(X,Y) = \frac{g^2}{2}\, {\rm Tr} \left[
\Gamma^\mu_a\, S(X,Y)\, \Gamma^\nu_b\, S(Y,X) \right]\;,
\ee
where the trace runs over Nambu-Gor'kov, color, flavor, and Dirac
indices. In momentum space, this expression reads
\be
\Pi^{\mu \nu}_{ab}(P) = \frac{g^2}{2} \, \int_K 
{\rm Tr} \left[ \Gamma^\mu_a\, S(K)\, \Gamma^\nu_b\, S(K-P) \right]\;.
\ee
For a two-flavor color superconductor, the gluon self-energy  has been
computed as a function
of energy $p_0$ and momentum ${\bf p}$ in Ref.\ \cite{DHR2flavor}. 
For other color-superconducting
phases, only the zero-energy, low-momentum limit of the 
gluon self-energy is known, from which one deduces the
Debye and Meissner masses. In the 
color-flavor-locked phase \cite{CFL} these were 
computed in Ref.\ \cite{DHRCFL}.
In the polar and color-spin-locked phases \cite{spinone}, where each quark
flavor pairs at its own Fermi surface and Cooper pairs carry total
spin one, the Debye and Meissner masses have been 
calculated in Ref.\ \cite{ASQWDHR}.

\section{Solution of the Yang-Mills equation} \label{IV}

In the following, we shall show that, in a two-flavor color superconductor,
the solution of the Yang-Mills equation (\ref{YM2})
is a constant background field 
$A_\mu^a \equiv A\, g_{\mu 0}\, \delta^{a8}$, $A= const.$.
As mentioned above, this background field acts like a color-chemical potential 
$\mu_8 \equiv g A$ for quarks. 

In order to solve the Yang-Mills equation (\ref{YM2}), we 
consider the source terms on the right-hand side.
Note first that the contribution from the last term in Eq.\ (\ref{YM2})
is absent in the mean-field approximation for $\Gamma_2$, Eq.\ 
(\ref{Gamma2}). Second, by dimensional arguments 
the first term in Eq.\ (\ref{YM2}), which
corresponds to the gluon tadpole contribution, is of order $\sim T^3$, 
and thus can be neglected for the temperature range of interest
for color superconductivity, $T \sim \phi 
\sim \mu \exp(-1/g) \ll \mu$ in weak coupling. Therefore, we only
need to consider the quark tadpole contribution which, on account
of Eq.\ (\ref{treequarkprop}), takes the form
\be \label{tadpole}
{\cal T}^\mu_a \equiv -\frac{1}{2}\,\frac{\delta }{\delta A_\mu^a(X)} \,
\int_{Y,Z} {\rm Tr} \left[ S_0^{-1}(Y,Z) S(Z,Y)\right] 
= - \frac{g}{2} \, {\rm Tr} \left[\Gamma^\mu_a \, S(0) \right]\; ,
\ee
where the trace only runs over Nambu-Gor'kov, color, flavor, and
Dirac indices. We used the fact that, for translationally invariant systems,
$S(X,Y) \equiv S(X-Y)$, i.e., $S(X,X) \equiv S(0)$. The tadpole contribution 
is simply a constant source term in the Yang-Mills
equation (\ref{YM2}) for the gluon field.

Taking $A_\mu^a \equiv 0$ in the quark propagator $S(0)$,
the tadpole contribution (\ref{tadpole}) has been computed for
a two-flavor color superconductor in Ref.\ \cite{Rebhan}, with the
result (\ref{Jmua}).
Let us try to find a solution of the Yang-Mills equation (\ref{YM2}),
with the expression (\ref{Jmua}) as a constant source term 
on the right-hand side. It is obvious
that the solution cannot be $A_\mu^a = 0$ for all $a= 1, \ldots , 8$. 
The perturbative calculation of the tadpole
performed in Ref.\ \cite{Rebhan}, with the result (\ref{Jmua}), 
is therefore not consistent
with the Yang-Mills equation (\ref{YM2}). For a self-consistent
solution, we have to allow for
a {\em non-zero expectation value of the gluon field\/} $A_\mu^a$ in
the full quark propagator $S(0)$. We have to repeat the calculation
of the tadpole with this non-zero background field, before we
plug the result back into the Yang-Mills equation (\ref{YM2}).

Let us first make a few general remarks concerning the structure of the
expectation value $A_\mu^a$. 
First, since we consider translationally invariant
systems, it has to be constant,
$A_\mu^a = const. \neq 0$. 
Furthermore, rotational symmetry requires $A_\mu^a \equiv g_{\mu 0} \, A^a$.
Denoting ${\cal T}^0_a \equiv {\cal T}_a$, 
we conclude
\be \label{YM4}
- g^2 \, f_{abc}\, f^{bde} \, A^c \, A_d\, A_e = {\cal T}_a \equiv 0 
\ee
by the antisymmetry of the structure constants.
Therefore, the expectation value of the 
gluon field in the quark propagator has to have precisely the
magnitude to {\em cancel the quark tadpole contribution}.

The calculation of the tadpole contribution {\em including\/}
a non-zero $A_\mu^a = g_{\mu 0}\, A^a$ is completely
analogous to the one performed in Ref.\ \cite{Rebhan} for
$A_\mu^a = 0$. The only difference is that one has to modify the
expressions for the
propagators of quasiparticle excitations with fundamental colors 1 and 2,
$G^\pm (Q)$ (Eq.\ (21) of Ref.\ \cite{Rebhan}), and for those of
unpaired quarks with fundamental color 3, $G_0^\pm(Q)$ (Eq.\ (22)
of Ref.\ \cite{Rebhan}). 
If we do not specify the color direction $a$ of the gluon field,
this is a non-trivial task, but we may simplify the calculation considerably
by the following argument.
Note first that, when
$A^a \equiv 0$ for all colors $a= 1, \ldots, 8$, only
the eighth color component of the tadpole is non-vanishing, cf.\
Eq.\ (\ref{Jmua}). Hence, we do not need to introduce non-zero
gluon fields for $a= 1, \ldots, 7$; the corresponding
tadpole contributions in these directions in adjoint color space
vanish already when $A^a =0$ for $a = 1, \ldots, 7$.
Obviously, we only need a non-vanishing component of the gluon field in the
eighth color direction, $A^a = A \, \delta^{a8}$, $A \neq 0$, which has
to be adjusted such that ${\cal T}_8 \equiv 0$.

As already mentioned in the introduction and
after Eq.\ (\ref{treequarkprop}),
a non-zero gluon field in the time direction, $A_\mu^a = A\, g_{\mu 0}\,
\delta^{a8}$, acts like
a color-chemical potential $\mu_8 \equiv g A$, which shifts the original
quark number chemical potential $\mu$.
For quarks of fundamental
colors 1 and 2 (which form Cooper pairs in a two-flavor
color superconductor) this shift is
\be \label{mu1}
\mu \rightarrow  \mu + g\, A^8_0 \, (T_8)_{11}
\equiv  \mu + g\, A^8_0 \,(T_8)_{22} 
= \mu + \frac{g A}{2 \sqrt{3}} \equiv \mu_1 \; ,
\ee
where we used $(T_8)_{11} = (T_8)_{22} = 1/(2 \sqrt{3})$,
and for quarks of fundamental color 3 (which remain unpaired
in a two-flavor color superconductor) it is
\be \label{mu2}
\mu \rightarrow \mu + g\, A^8_0 \,(T_8)_{33} = 
\mu - \frac{g A}{\sqrt{3}} \equiv \mu_2\;,
\ee
where we used $(T_8)_{33} = - 1/\sqrt{3}$.
In the evaluation of the tadpole contribution by Gerhold and Rebhan
\cite{Rebhan} we now
have to replace $\mu$ by $\mu_1$ in their Eq.\ (21) for the propagator
$G^\pm(Q)$ (which describes the propagation of quasiparticles
with fundamental colors 1 and 2),
\be
G^\pm (Q) = \sum_{e=\pm} \frac{ q_0 \mp (\mu_1 - eq)}{
q_0^2 - (\mu_1 - e q)^2 - |\phi^e|^2} \, \Lambda^{\pm e}({\bf q}) \gamma_0 \; .
\ee
Likewise, we have to replace
$\mu$ by $\mu_2$ in their Eq.\ (22) for the propagator
$G_0^{\pm}$ (which describes the propagation of unpaired
quarks with fundamental color 3),
\be
G_0^\pm (Q) = \sum_{e=\pm} \frac{ q_0 \mp (\mu_2 - eq)}{
q_0^2 - (\mu_2 - e q)^2 } \, \Lambda^{\pm e}({\bf q}) \gamma_0 \; .
\ee
Then, their calculation can be copied up to Eq.\ (24), which now reads
\be \label{newtad}
{\cal T}_8 = \frac{g }{\pi^2}\, (T_8)_{33} \int_0^\infty {\rm d} q\, q^2
\left( \frac{\mu_1 - q}{\sqrt{(\mu_1 - q)^2 + \phi^2}}
     - \frac{\mu_2 - q}{|\mu_2 - q |} \right)\,\, .
\ee
(We take the gap function $\phi$ to be real-valued.)
Some care has to be taken in the evaluation of this integral. 
The difference in the chemical potentials, 
$\delta \mu \equiv \mu_2 - \mu_1 \equiv
- \sqrt{3}g A/2$ has to be of the right magnitude
to achieve the cancellation of the tadpole, ${\cal T}_8 \equiv 0$. 
As will be shown below, $\delta \mu \sim \phi^2/(g \mu)$, i.e.,
it is parametrically much smaller than the gap.
For the following calculation, it is therefore permissible to make
the assumption $\delta \mu \ll \phi$, which is consistent with our final
result. We split the integral into three regions, 
(I) $0 \leq q < \mu_1 - \delta$, (II)
$\mu_1 - \delta \leq q \leq \mu_1 + \delta$, 
and (III) $\mu_1 + \delta < q < \infty$. The quantity
$\delta$ is chosen such that the gap function $\phi$, which
is known to rapidly vanish away from the Fermi surface
\cite{gapequation}, can be neglected in regions (I) and (III).
We do not need to specify the magnitude of $\delta$, it
suffices to know that $\phi \ll \delta \ll \mu$.
We may then write
\bea 
{\cal T}_8 & \simeq &
\frac{g}{\pi^2}\, (T_8)_{33} \left\{
\int_{\mu_1 - \delta}^{\mu_1 + \delta} {\rm d} q\, q^2
\left( \frac{\mu_1 - q}{\sqrt{(\mu_1 - q)^2 + \phi^2}}
     - \frac{\mu_2 - q}{|\mu_2 - q |} \right) \right. \nonumber \\
&   & \hspace*{1.3cm} +\left. \left[ \int_{0}^{\mu_1 - \delta} 
+ \int_{\mu_1 + \delta}^\infty \right] {\rm d} q\, q^2 
\left[ {\rm sign}(\mu_1 - q) - {\rm sign}(\mu_2 - q) \right]
\right\}
\,\, . \label{newtad2}
\eea
Since $\delta \gg \phi \gg \delta \mu$, the second integral always
vanishes. The remaining integral may be evaluated by substituting
$\xi \equiv q - \mu_1$. The integral over the first term
in the integrand is
\be \label{1}
- \int_{ - \delta}^{ \delta} {\rm d} \xi\, (\mu_1 + \xi)^2
 \frac{\xi}{\sqrt{\xi^2 + \phi^2}} \simeq - 2 \mu_1 \left(\delta^2
- \phi^2 \ln \frac{ 2 \, \delta}{\phi} \right)\;,
\ee
where we only kept the leading and subleading terms and assumed
the gap function to be constant over the range of integration.
The integral over the second term is best evaluated by shifting
the integration variable $\xi \rightarrow \xi' = \xi - \delta \mu$,
\be \label{2}
\int_{- \delta - \delta \mu}^{\delta- \delta \mu} {\rm d} \xi' 
(\mu_2 + \xi')^2 \; {\rm sign} \,\xi'
\simeq 2 \mu_2 \left( \delta^2 - \mu_2 \, \delta \mu \right)\;,
\ee
where we also only kept the leading and subleading terms.
Inserting the results (\ref{1}) and (\ref{2}) into Eq.\ (\ref{newtad2}) we
obtain
\be
{\cal T}_8 \simeq \frac{2\, g}{ \pi^2}\, (T_8)_{33} 
\left[  \mu_1 \, \phi^2 \ln \frac{ 2 \, \delta}{\phi}  - 
\delta  \mu \left(\mu_2^2 - \delta^2\right) \right]
\simeq  \frac{2 \,g }{ \pi^2}\, (T_8)_{33}\, \mu^2 \left( 
\frac{\phi^2 }{\mu} \ln \frac{2\, \delta}{\phi}  - \delta \mu \right) \;,
\ee
where in the last step we approximated $\mu_1 \simeq \mu_2 \simeq \mu$
and neglected higher-order terms, $\delta^2 \ll \mu^2$. 
To leading order, we may also assume that $2\, \delta / \phi \simeq
\exp[ 3 \pi^2/(\sqrt{2} g)]$ \cite{gapequation}. 
Consequently, the tadpole vanishes if
\be
\delta \mu \equiv - \frac{\sqrt{3}}{2}\, g A \simeq \frac{3 \pi^2}{\sqrt{2}}\,
\frac{\phi^2}{g\mu} \left[ 1 + O(g)\right]\;.
\ee
This results justifies our original assumption $\delta \mu \ll \phi$.
The expectation value of the gluon field necessary to ensure
a vanishing tadpole contribution is thus
\be \label{A8}
A_\mu^8 \equiv A\, g_{\mu0}\, \delta^{a8}\;\; ,
\;\;\;\; A \simeq - \sqrt{6}\, \pi^2 
\, \frac{\phi^2}{g^2 \mu} \left[ 1 + O(g) \right] \;.
\ee
As claimed above, this gives rise to
a chemical potential $\mu_8 \equiv g A$ of precisely
the magnitude (\ref{mu8}). 
In a self-consistent
calculation, it arises from a non-vanishing expectation value (\ref{A8}) of the
gluon field $A_\mu^a$, which has precisely the magnitude 
to cancel the tadpole. Let us mention that this ``tadpole
cancellation'' is a well-known mechanism in quantum field theories
with spontaneously broken symmetries: upon resummation,
perturbatively non-vanishing
tadpole contributions like (\ref{Jmua}) induce non-vanishing
expectation values for one-point functions which ultimately
cancel the tadpoles. Note also that $A < 0$, so that 
the Fermi surface for red and green quarks is {\em reduced},
cf.\ Eq.\ (\ref{mu1}),
while the one for blue quarks is {\em increased}, cf.\ Eq.\ (\ref{mu2}).
As argued in the introduction, this works in a way to
ensure color neutrality, cf.\ also next section.

\section{Color neutrality} \label{V}
 
Let us compute the color-charge density $n_a$ associated with adjoint
color $a$. To this end, introduce an {\em explicit\/} 
color-chemical potential $\mu_a$
into the inverse tree-level quark propagator (\ref{treequarkprop}). 
This amounts to adding a term $\mu_a \Gamma^0_a$ to
$S_0^{-1}$, where $\Gamma^0_a$ is given by Eq.\ (\ref{NGvertex}).
The thermodynamic pressure 
is $p\equiv (T/V) \Gamma^*$, where $\Gamma^*$ 
is the effective action (\ref{Gamma}), evaluated at
the stationary point defined by Eqs.\ (\ref{Dirac}) -- (\ref{DSEgluon}).
The color-charge density $n_a$ is then given by
\be \label{na}
n_a \equiv \frac{\partial p}{\partial \mu_a}
= \frac{1}{2} \, \frac{T}{V} \int_{X,Y} {\rm Tr} \left[ S(X,Y)\,
\frac{\partial S_0^{-1} (Y,X)}{\partial \mu_a} \right]
= \frac{1}{2} \, {\rm Tr} \left[ \Gamma^0_a \, S(0) \right]\;,
\ee
where the traces run only over Nambu-Gor'kov, color, flavor, and Dirac
indices. Other terms cancel after making use of
Eqs.\ (\ref{DSE2}) and (\ref{Sigma}). Finally, we set
$\mu_a = 0$ to compute the color-charge density for {\em vanishing\/}
color-chemical potential. If this were non-zero, 
color superconductivity would break the
color gauge symmetry not only spontaneously, but also {\em explicitly}
by generating a (net) color charge $n_a$.
This must not be the case. Comparing Eqs.\ (\ref{na})
and (\ref{tadpole}), we obtain the identity
\be
n_a \equiv - \frac{1}{g} \, {\cal T}_a^0 \equiv - \frac{1}{g} \, {\cal T}_a
\;.
\ee
The {\em cancellation of the tadpole contribution\/}, ${\cal T}_a = 0$,
by a {\em non-vanishing expectation value\/} $A_\mu^a = A \, g_{\mu 0}\,
\delta^{a8}$ for the gluon field, as demonstrated
in Sec.\ \ref{IV}, thus also ensures {\em color neutrality}.
This result also implies that, in QCD, color is not broken
explicitly by the condensation of Cooper pairs.

\section*{Acknowledgement}
D.D.D.\ acknowledges support by GSI Darmstadt, BMBF, and DAAD.
D.H.R.\ thanks J.\ Bowers and T.\ Sch\"afer for discussions, as well as
A.\ Gerhold, A.\ Rebhan, I.\ Shovkovy, and Q.\ Wang for a critical
reading of the manuscript. 
He also acknowledges support by the INT Seattle where part of this work
was done.

\end{document}